\documentclass[preprint,12pt]{elsarticle}

\usepackage{amssymb}
\usepackage{amsmath}
\usepackage{multicol}
\usepackage{natbib}

\newlength{\onecol}
\newlength{\twocol}
\newcommand{\ipi}{{i-PI}}
\newcounter{bla}

\journal{Computer Physics Communications}

\begin{document}

\begin{frontmatter}




\title{ \ipi{}: A Python interface for ab initio path integral molecular dynamics simulations}

\author[a]{Michele Ceriotti\corref{author}}
\author[a]{Joshua More}
\author[a]{David E. Manolopoulos}

\cortext[author] {Corresponding author.\\\textit{E-mail address:} michele.ceriotti@chem.ox.ac.uk}
\address[a]{Physical and Theoretical Chemistry Laboratory, 
University of Oxford, South Parks Road, Oxford OX1 3QZ, UK}

\begin{abstract}
Recent developments in path integral methodology have significantly reduced the
computational expense of including quantum mechanical effects in the nuclear motion
in \emph{ab initio} molecular dynamics simulations. However, the 
implementation of these developments requires a considerable programming 
effort, which has hindered their adoption. 
Here we describe \ipi{}, an interface written in Python that has been designed to minimise
the effort required to bring state-of-the-art path integral techniques to an 
electronic structure program.  While it is best suited to first principles calculations
and path integral molecular dynamics, \ipi{} can also be used to perform classical
molecular dynamics simulations, and can just as easily be interfaced with an empirical
forcefield code. To give just one example of the many potential applications
of the interface, we use it in conjunction with the CP2K electronic structure package to 
showcase the importance of nuclear quantum effects in high pressure water.

\end{abstract}


\end{frontmatter}


\section{Introduction}

Molecular dynamics simulations are becoming increasingly capable not only of
assisting the interpretation of experiments, but also of predicting the 
properties and the behaviour of new materials and compounds~\cite{fisc+06natm}.
Besides the increase in available computer power, these developments have been
made possible by an increasingly accurate treatment of the interactions
between the atoms, in particular by an explicit treatment
of the electronic structure problem~\cite{car-parr85prl,burk12jcp}. 
However, as the methods that are used to treat the quantum nature of the electrons improve, it
becomes increasingly clear that in the presence of light atoms, such as
hydrogen or lithium, the error due to approximating the nuclei as 
classical particles is at least as large as the errors due to the 
approximate modelling of the electrons. The importance
of nuclear quantum effects (NQEs) is evident from the fact that the zero-point 
energy associated with a typical O--H stretching mode is in excess of 200~meV.
This has significant implications: for example, one can show by extrapolating the experimental values
for isotopically pure light, heavy and tritiated water that a hypothetical
liquid with classical nuclei would have a 50\% higher heat capacity than H$_2$O
and a pH of about 8.5.

Within the Born-Oppenheimer approximation, NQEs can be modelled 
using the imaginary time path integral 
formalism~\cite{chan-woly81jcp,parr-rahm84jcp,cepe95rmp,feyn-hibb65book}.
This formalism maps the quantum mechanical partition 
function for a set of distinguishable nuclei onto the classical
partition function of a so-called ring polymer, composed of $n$ replicas (beads)
of the physical system connected by harmonic springs. 
The number of replicas needed to achieve a converged result is typically
a small multiple of $\beta\hbar\omega_{\rm max}$, where $\beta$ is 
the inverse temperature and $\omega_{\rm max}$ is the largest vibrational
frequency in the system. For the archetypical case of room-temperature 
liquid water, most properties are reasonably well converged with $n=32$ -- making 
the path integral simulation 32 times more expensive than a classical
simulation. 

As a consequence of this large overhead, NQEs were for many years
only rarely considered in the context 
of {\emph{ab initio}} molecular dynamics~\cite{tuck+96jcp,marx-parr96jcp,marx+99nat}.
However, with the advent of massively parallel computers,
including these effects has become somewhat more affordable \cite{haye+09jpcb,chen+13pccp}, 
and their simulation has also been facilitated by new methodological developments.
In particular, it has recently been realised that an approximate modelling of NQEs can be obtained
by applying a coloured (correlated)-noise Langevin equation thermostat to classical molecular
dynamics~\cite{buyu+08pre,ceri+09prl2,damm+09prl}. Moreover,
this idea can be turned into a systematically convergent, accurate 
method~\cite{ceri+11jcp,ceri-mano12prl} by combining
coloured noise with path integral molecular dynamics (PIMD).
To give an example, the PIGLET method~\cite{ceri-mano12prl} makes it possible to treat NQEs in liquid 
water at 300~K with as few as 6 beads, which promises to make
accurate \emph{ab initio} PIMD calculations almost routine. 

Unfortunately, most \emph{ab initio} electronic structure codes only contain
rudimentary implementations of PIMD, if any at 
all. We have developed \ipi{} in order to reduce the effort of introducing the latest 
PIMD developments into electronic structure codes, so as to make them readily available to a broader
community. The framework we have adopted is a server-client model, in which \ipi{}
acts as a server passing atomic coordinates to (multiple instances of) an electronic structure
client program, and receiving the energy and forces in return. 
This allows all of the PIMD machinery to be confined to the Python server side, 
and minimises the modifications that have to be made to the electronic structure code.

\section{Program overview}

\ipi{} has been developed with the awareness that the cost of an \emph{ab initio}
PIMD simulation will be dominated by the electronic structure problem, and with
a few clear goals:
\begin{itemize}
\item to minimise the effort required to modify the client \emph{ab initio} electronic 
structure code,
\item to include no feature specific to a particular client code,
\item to be modular, simple to extend and with a structure that reflects the
underlying physics,
\item to be efficient, but never at the cost of clarity.
\end{itemize}

\begin{figure}[hpbt]
\centering\includegraphics[width=\twocol]{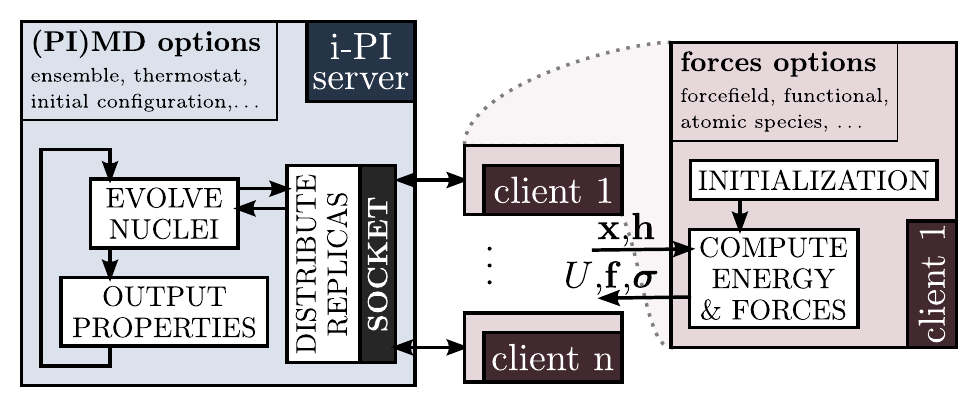}
\caption{\label{fig:scheme} Schematic overview of the client-server model underlying \ipi{}. 
The communication is kept to a minimum, and so are the modifications that need to be
made to adapt an existing electronic structure code to act as the client.
The client is not restarted between successive force evaluations, so that the 
overhead associated with initialisation is avoided.
}
\end{figure}

In order to acheive these goals we have chosen the Python programming language, and
a client-server model in which \ipi{} acts as the server, dealing exclusively with 
evolving the nuclear degrees of freedom according to the equations of motion. 
The evaluation of the forces, the potential and the virial is delegated to one or more 
instances of the client code (see Figure~\ref{fig:scheme}). 
\ipi{} maintains a list of active clients, to which it 
dispatches the positions of the nuclei and from which it collects
the forces and energy as soon as they have been computed. The communication is kept
to a minimum, not so much for its impact on performance, as for the fact 
that exchanging more information would require a more substantial implementation effort 
on the client side.
All the details of the force evaluation -- such as the parameters of the electronic
structure calculation -- are left to the input of the external code: \ipi{} only sends the 
dimensions of the simulation box ($\mathbf{h}$) and the coordinates of the nuclei ($\mathbf{x}$), 
and expects the 
atomic species to be stored internally by the client code in the same order they are 
in the \ipi{} input. The client code computes and returns the ionic forces ($\mathbf{f}$), 
electronic energy ($U$) and stress tensor ($\boldsymbol{\sigma}$), 
possibly supplemented by a string that contains any further information
on the electronic structure (e.g. the dipole moment, the partial charges of the atoms, etc.)
that might be required for post-processing; \ipi{} simply outputs this string verbatim.

The client-server communication is implemented using sockets. 
If desired, \ipi{} can communicate with the clients across the internet, 
realising a rudimentary distributed computing paradigm.
If one wishes to reduce the communication overhead -- e.g. if one wants to use \ipi{}
with an empirical forcefield driver -- local UNIX-domain sockets can be used.
It would also be easy to implement different communication methods, e.g. via inter-process
MPI or writing to disk. 

Instances of the patched \emph{ab initio} code can register themselves dynamically, 
and \ipi{} keeps track of the active clients so that the forces on multiple path integral replicas
can be evaluated at once. The trivial layer of parallelism over the
beads can therefore be fully exploited. We would recommend  implementing 
the communication code into the client software in a way that mimics one
of the existing procedures that update nuclear positions, as is done in molecular 
dynamics,  geometry optimisation, etc. 
The electronic structure infrastructure only has 
to be initialised once, before the client connects to the \ipi{}
server and waits to receive a set of atomic coordinates.
It then performs an electronic structure calculation, returns
the energy, forces and stress tensor to the server, and waits
for the next set of coordinates. 
 The converged density/wavefunction from the previous step
can be re-used as the starting point for the next, which is advantageous because
 in all probability this will correspond to a very similar atomic configuration. 
 This makes our approach superior to a scripting approach in
which the client code is re-launched with a new input for each 
atomic configuration.
In order to fully take advantage of the density/wavefunction
extrapolation techniques implemented in many modern electronic structure codes, \ipi{}
will always try to dispatch the same replica to a given instance of 
the client, to ensure that a smooth sequence of configurations is provided 
to each client from one time step to the next.

\begin{figure}[hpbt]
\centering\includegraphics[width=\twocol]{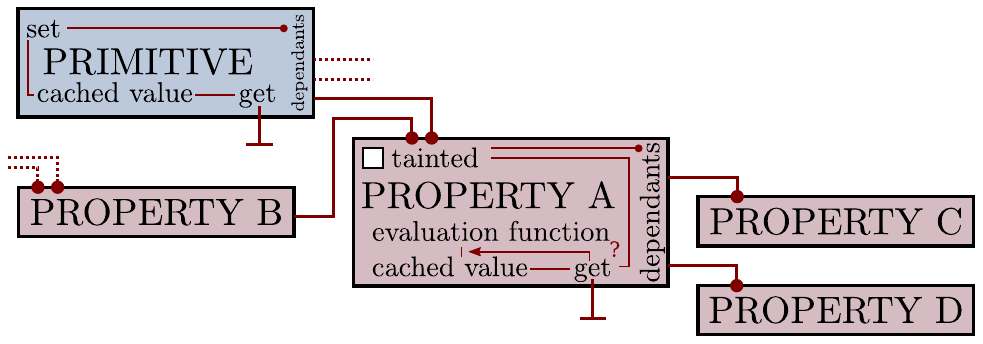}
\caption{\label{fig:depend} Schematic overview of the functioning of the 
\emph{depend} class used as the base for properties and physical quantities in \ipi{}. 
A few ``primitive'' quantities -- such as atomic positions or momenta -- can be modified
directly. For most properties, one defines a function that can compute their value 
based on the values of other properties. Whenever one property is modified, all the quantities that
depend on it are marked as tainted, so that -- when the value of one of the properties
is used, the function can be invoked and the updated value can be obtained and 
cached. If the same quantity is requested again and it has not been 
marked as tainted in the mean time, the cached value is returned. 
}
\end{figure}

To make the internal coding of \ipi{} as transparent and as close to the physics 
of the problem as possible, we have introduced a dedicated Python class
to represent physical quantities -- such as the atomic coordinates, the momenta, the kinetic
energy, etc. Instances of this class can be used simply as variables containing
the value of the quantity they represent. However, they also contain information on 
how each quantity depends on others, and how it can 
be computed from its dependencies (see Figure~\ref{fig:depend}). 
Derived quantities are computed automatically when requested, 
and their value is cached for future reference. If the same quantity is 
needed again, the cached value will be used, avoiding the overhead of 
re-computing it. If however one of the quantities it 
depends on has been modified, it will be re-evaluated automatically.
This mechanism of dependency detection and value caching considerably de-clutters
the code, since one does not need to keep track of when quantities need to be 
recomputed. 

The input for \ipi{} consists in a xml file, containing a number of 
tags and options that specify the details of the simulation. The list of 
available options are described in detail in the manual. \ipi{} can dump
the full state of the simulation as \emph{checkpoint} files, that can 
be used to seamlessly re-start a simulation and that have the same format
of an input file, so it is easy to modify some of the options, if desired.

\section{Program features}

Integrating the PIMD equations of motion involves a number
of technicalities. Fortunately, these can be kept to a minimum by using 
stochastic thermostatting~\cite{ceri+10jcp} -- which avoids the  
complication of integrating Nos{\'e}-Hoover chain thermostats -- and by
formulating the time evolution using a symmetric Trotter splitting 
algorithm~\cite{tuck+92jcp}. This is the approach we have adopted in the \ipi{} program.

The program implements the most recent developments in path integral and
colored-noise molecular dynamics, including the following:
\begin{itemize}
\item molecular dynamics and PIMD in the {\em NVE}, {\em NVT} and {\em NPT} ensembles, 
with the high-frequency internal vibrations of the path propagated in the
normal-mode representation~\cite{ceri+10jcp};
\item ring polymer contraction~\cite{mark-mano08jcp,mark-mano08cpl}, implemented 
by exposing multiple socket interfaces to deal separately with short and long-range
components of the potential energy;
\item efficient stochastic velocity rescaling \cite{buss+07jcp} and 
path integral Langevin equation thermostats~\cite{ceri+10jcp};
\item various generalized Langevin equation (GLE) thermostats,
including the optimal sampling~\cite{ceri+09prl,ceri+10jctc}, quantum 
~\cite{ceri+09prl2}, and $\delta$~\cite{ceri-parr10pcs} thermostats, the parameters 
for which can be downloaded from an on-line repository~\cite{gle4md};
\item mixed path integral--generalized Langevin equation techniques for 
accelerated convergence, including both PI+GLE~\cite{ceri+11jcp} and the 
more recent and effective version PIGLET~\cite{ceri-mano12prl};
\item all the standard estimators for structural properties, the quantum 
kinetic energy, pressure, etc.;
\item more sophisticated estimators such as the scaled-coordinate 
heat capacity estimator~\cite{yama05jcp}, 
estimators to obtain isotope fractionation free energies by re-weighting a 
simulation of the most abundant isotope~\cite{ceri-mark13jcp}, and 
a displaced-path estimator for the particle momentum distribution~\cite{lin+10prl};
\item the infrastructure that is needed to perform ring polymer molecular dynamics 
(RPMD)~\cite{crai-mano04jcp,habe+13arpc} and centroid molecular dynamics
(CMD)~\cite{cao-voth93jcp,cao-voth94jcp} approximate quantum dynamics calculations.
\end{itemize}

\section{Constant-pressure path integral molecular dynamics}

Most of the techniques listed above have been discussed in detail elsewhere. However,
our implementation of PIMD in the {\em NPT} ensemble has not been described before now. The
approach we have adopted for this is a relatively straightforward combination of ideas taken from
path integral Langevin equation thermostats~\cite{ceri+10jcp}, stochastic barostats 
for conventional MD~\cite{buss+09jcp}, and previous work on constant-pressure 
PIMD~\cite{mart+99jcp}. We report it here as this combination is robust, transparent
and streamlined, and we think it could be useful as a reference for future implementations of 
PIMD and PIGLET in the {\em NPT} ensemble.

Consider a classical system composed of $N$ atoms, described by the Hamiltonian
\begin{equation}
H({\bf p},{\bf q}) = \sum_{i=1}^N \frac{{\bf p}_i^2}{2m_i} +U({\bf q}_1,\ldots, {\bf q}_N), 
\end{equation}
where $U$ is the potential energy of the system, and $m_i$ the mass of the $i$-th atom. 
The path integral Hamiltonian for the same system reads~\cite{ceri+10jcp}
\begin{equation}
H_n({\bf p},{\bf q}) = H_n^0({\bf p},{\bf q})+ \sum_{j=1}^n U\left({\bf q}^{(j)}_1,\ldots, {\bf q}^{(j)}_N\right), 
\end{equation}
where ${\bf q}_i^{(j)}$ contains the Cartesian coordinates of the $i$-th atom in the $j$-th 
replica and 
\begin{equation}
H_n^0({\bf p},{\bf q}) = \sum_{i=1}^{N}\sum_{j=1}^n \left(\frac{|{\bf p}_{i}^{(j)}|^2}{ 2m_i}+
\frac{1}{2}m_i\omega_n^2|{\bf q}_{i}^{(j)}-{\bf q}_{i}^{(j-1)}|^2\right), 
\end{equation}
is the free-particle ring-polymer Hamiltonian. Here 
$\omega_n=n/\beta\hbar$ and cyclic boundary 
conditions are implied: ${\bf q}_{i}^{(j)}\equiv {\bf q}_{i}^{(j+n)}$.

We have chosen to implement a version of the {\em NPT} ensemble in which only the centroid coordinates $\bar{\bf q}_i=\frac{1}{n}\sum_j {\bf q}_{i}^{(j)}$ are scaled when the simulation 
cell volume $V$ is changed~\cite{mart+99jcp}. Following previous literature on 
molecular dynamics at constant pressure~\cite{ande80jcp,parr-rahm81jap,mart+99jcp,buss+09jcp}
we allow the cell volume to fluctuate, assigning a fictitious \lq mass' $\mu$ to the cell 
and using the log-derivative of the cell volume $V$ with respect to time to define the cell \lq momentum'
$\alpha=\mu \dot{V}/3V$. The fictitious mass can be related to a characteristic 
relaxation time scale for the cell dynamics, $\tau_\alpha$, by $\mu=3N\tau_\alpha^2/\beta$. 

Let us start by stating the equations of motion we will
use; we will then move on to describe the corresponding stationary distribution, and finally how
the equations of motion can be integrated. Although other stochastic thermostats could  
be used, we will consider here the case of a white-noise Langevin
piston and of a configurational PILE-L thermostat on the ring polymer normal modes~\cite{ceri+10jcp}.
We will write the equations of motion in the normal mode representation,
using $\tilde{\mathbf{q}}_{i}^{(k)}=\sum_j {\mathbf{q}}_{i}^{(j)} C_{jk}$ 
(where \(C_{jk}\) is defined in Ref.~\cite{ceri+10jcp}) to indicate the 
$k$-th normal-mode component corresponding to the quantity $\mathbf{q}$ for the $i$-th atom, and 
$\tilde{\mathbf{f}}_i^{(k)}$ as a short-hand to indicate the force $-\partial U/\partial\tilde{\mathbf{q}}_i^{(k)}$. 
In this notation, equations of motion for ${\tilde{\mathbf{p}}}^{(k)}_i$ and
${\tilde{\mathbf{q}}}^{(k)}_i$, split according to the same scheme we will
use for the Liouville operator further down, are as follows:
\begin{subequations}
\begin{align}
\dot{\tilde{\mathbf{p}}}^{(k)}_i = & \sqrt{\frac{2nm_i\gamma_k}{\beta}} \boldsymbol{\xi}^{(k)}_i 
-\gamma_k {\tilde{\mathbf{p}}}^{(k)}_i \label{eq:motion-gamma} \\
\label{eq:motion-force}
 &+ {\tilde{\mathbf{f}}^{(k)}_i } \\
\label{eq:motion-pscale}
&- \tilde{\mathbf{p}}^{(k)}_i \delta_{k0}\, {\alpha}/{\mu} \\
\label{eq:motion-nm}
\begin{aligned}
\ \\
\dot {\tilde{\mathbf{q}}}^{(k)}_i =
\end{aligned}&
\left.\begin{aligned}
&- m_{i}\omega_k^2 \tilde{\mathbf{q}}^{(k)}_i \\
& {\tilde{\mathbf{p}}^{(k)}_i }/{m_i} \\
\end{aligned} \quad\quad \right\}  \\ 
\label{eq:motion-qscale}
 & + \tilde{\mathbf{q}}^{(k)}_i  \delta_{k0} \, {\alpha}/{\mu} \\
\label{eq:motion-volume}
\dot {V} = & 3 V \,{\alpha}/{\mu} \\
\label{eq:motion-gamma-alpha}
\dot{\alpha} = & \sqrt{\frac{2 n\mu \gamma_\alpha}{\beta}} \xi_\alpha - \gamma_\alpha \alpha 
\\
\label{eq:motion-alpha}
  & + 3 n \left[V \left(P_{\rm int}-P_{\rm ext}\right)+\frac{1}{\beta}\right].
\end{align}
\end{subequations}
Here $\xi_{\alpha}$ is a scalar and $\boldsymbol{\xi}^{(k)}_i$ a vector of uncorrelated 
normal deviates with zero mean and unit 
variance, and we use the centroid-virial estimator for the internal pressure in the form
\begin{equation}\label{eq:pressure}
P_\mathrm{int}=-\frac{1}{n}
\sum_j\frac{\mathrm{d}U(\mathbf{q}^{(j)})}{\mathrm{d}V} +
\frac{1}{3nV}\left[\sum_i \frac{|\tilde{\mathbf{p}}_i^{(0)}|^2}{m_i} + \sum_{ij} \left(\mathbf{q}_i^{(j)}-\bar{\mathbf{q}}_i\right)\cdot \frac{\partial U(\mathbf{q}^{(j)})}{\partial \mathbf{q}_i^{(j)}}
\right].
\end{equation}
Note that the kinetic energy term in this estimator could equally well be substituted by its mean,
$\left<\sum_i {|\tilde{\mathbf{p}}_i^{(0)}|^2}/{m_i} \right>=3Nn/\beta$. However,  
we use the instantaneous value as this is needed for the dynamics to have a 
well-defined conserved quantity. Note also that we use ${\mathrm{d}U(\mathbf{q}^{(j)})}/{\mathrm{d}V}$ 
to signify the \emph{total} derivative of the potential energy with respect to volume, which typically 
contains a term obtained from the virial. 
We shall not enter into the details of how the virial should be computed, 
since the evaluation of the potential energy 
component of the pressure is delegated to the client program. Regardless of the implementation,
one should make sure that the client returns the total derivative, including terms that depend explicitly on
the volume such as tail corrections.

It is straightforward but tedious to show that the equations of motion in Eqs.~\eqref{eq:motion-gamma}-\eqref{eq:motion-alpha}
have the conserved quantity
\begin{equation}
E_\mathrm{cons}=H_n({\bf p},{\bf q})+n P_\mathrm{ext} V+\frac{\alpha^2}{2\mu} - \frac{n}{\beta}\ln V+\Delta H,
\end{equation}
where $\Delta H$ is the heat transfer balance of the thermostatting steps in Eqs.~\eqref{eq:motion-gamma}
and~\eqref{eq:motion-gamma-alpha}, as explained in Refs.~\cite{buss-parr07pre,ceri+10jcp}.

Similarly, one can show that the probability distribution
\begin{equation}
 \rho({\bf p},{\bf q},\alpha,V) \propto \exp\left(-\frac{\beta}{n}\left[H_n({\bf p},{\bf q})+n P_\mathrm{ext} V+\frac{\alpha^2}{2\mu}
\right] \right)\label{eq:measure}
\end{equation}
is stationary with respect to the Liouville operator
\begin{align}
\begin{split}
\mathcal{L}=&\sum_{ik}\mathcal{L}_\gamma^{(ik)}
           + \mathcal{L}_{\gamma_\alpha}
           +\sum_{ik}\mathcal{L}_U^{(ik)}     
           +\mathcal{L}_\alpha \\     
           &+\left(\sum_{i}\mathcal{L}_0^{(i0)} 
             +\sum_{i}\mathcal{L}_p^{(i)}
             +\sum_{i}\mathcal{L}_q^{(i)}
              + \mathcal{L}_V\right)
              +\sum_{i(k>0)}\mathcal{L}_0^{(ik)}
              \\
          =&\ \mathcal{L}_\gamma
          +\mathcal{L}_{\gamma_\alpha}
          +\mathcal{L}_U
          +\mathcal{L}_\alpha
          +\mathcal{L}_0^0
          +\mathcal{L}_0' ,
\end{split}
\label{eq:liouville}
\end{align}
where the splitting of the Liouville operator is that arising from the
subdivision of the equations of motion~\eqref{eq:motion-gamma}-\eqref{eq:motion-alpha}:
\begin{subequations}
\begin{align}  
\label{eq:liouville-gamma}
   \mathcal{L}_\gamma^{(ik)} = & -\gamma_k
  \frac{\partial ( \tilde{\mathbf{p}}_i^{(k)}\cdot)}{\partial \tilde{\mathbf{p}}_i^{(k)} } 
 - \frac{nm_i\gamma_k}{\beta} \frac{\partial^2\,\cdot}{\partial {\tilde{\mathbf{p}}_i^{(k)2}}}   \\
\label{eq:liouville-force}
  \mathcal{L}_U^{(ik)} = & \tilde{\mathbf{f}}_i^{(k)} \frac{\partial \,\cdot}{\partial \tilde{\mathbf{p}}_i^{(k)} } \\
\label{eq:liouville-pscale}
   \mathcal{L}_{p}^{(i)} = & -\frac{\alpha}{\mu} \frac{\partial (\tilde{\mathbf{p}}_i^{(0)}\cdot)}{\partial \tilde{\mathbf{p}}_i^{(0)} } \\
\label{eq:liouville-nm}
\mathcal{L}_0^{(ik)}= & \frac{\tilde{\mathbf{p}}^{(k)}_i }{m_i} \frac{\partial \,\cdot}{\partial \tilde{\mathbf{q}}_i^{(k)} }
-m_{i}\omega_k^2 \tilde{\mathbf{q}}^{(k)}_i \frac{\partial\,\cdot }{\partial \tilde{\mathbf{p}}_i^{(k)} } \\ 
\label{eq:liouville-qscale}
  \mathcal{L}_{q}^{(i)}= & \frac{\alpha}{\mu} \frac{\partial (\tilde{\mathbf{q}}_i^{(0)} \cdot)}{\partial \tilde{\mathbf{q}}_i^{(0)} }  \\
\label{eq:liouville-volume}
\mathcal{L}_V = & 3\frac{\alpha}{\mu}\frac{\partial (V\cdot)}{\partial V} \\
\label{eq:liouville-gamma-alpha}
\mathcal{L}_{\gamma_\alpha} = &-\gamma_\alpha \frac{\partial(\alpha \cdot)}{\partial \alpha} -
\frac{n\mu\gamma_\alpha}{\beta} \frac{\partial^2 \,\cdot}{\partial\alpha^2}
\\
\label{eq:liouville-alpha}
\mathcal{L}_\alpha = & 3 n \left[V \left(P_{\rm int}-P_{\rm ext}\right)+\frac{1}{\beta}\right] \frac{\partial \,\cdot}{\partial\alpha}.
\end{align}
\end{subequations}

Note that we have modified slightly the equations of motion given in~\cite{buss+09jcp} by including just 
one $1/\beta$ in Eq.~\eqref{eq:motion-alpha} and~\eqref{eq:liouville-alpha} rather than two, so that 
we obtain the same {\em NPT} 
stationary distribution as in Ref.~\cite{mart+99jcp} (see Eq.~\eqref{eq:measure}). 
Also, the $P_{\rm ext}V$ term in this distribution is multiplied by the number of beads, which is 
consistent with the fact that the path integral partition function is evaluated at
$n$ times the physical temperature.

The integration scheme we have implemented in \ipi{} follows closely that of Ref.~\cite{buss+09jcp}
and is based on the following symmetric Trotter splitting of the propagator
\begin{equation}
\begin{split}
e^{-\mathcal{L}\Delta t}\approx &
e^{-(\mathcal{L}_\gamma+\mathcal{L}_{\gamma_\alpha})\Delta t/2}
e^{-(\mathcal{L}_U+\mathcal{L}_\alpha)\Delta t/2}   
e^{-(\mathcal{L}_0^0+\mathcal{L}_0')\Delta t} 
e^{-(\mathcal{L}_U+\mathcal{L}_\alpha)\Delta t/2}
e^{-(\mathcal{L}_\gamma+\mathcal{L}_{\gamma_\alpha})\Delta t/2}.
\end{split}\label{eq:bigliouville}
\end{equation}
Thus, the thermostat is first applied to the ring polymer and cell momenta
for half a time step (Eqs.~\eqref{eq:motion-gamma} and~\eqref{eq:motion-gamma-alpha}):
\begin{equation}
\begin{split}
\alpha \,\leftarrow\, &  
e^{-\gamma_\alpha\Delta t/2} \alpha + \sqrt{\frac{n\mu}{\beta}(1-e^{-\gamma_\alpha\Delta t})} \xi_\alpha 
\\
\tilde{\mathbf{p}}_i^{(k)} \,\leftarrow\, &
 e^{-\gamma_k\Delta t/2} \tilde{\mathbf{p}}_i^{(k)}
   + \sqrt{\frac{nm_i}{\beta}(1-e^{-\gamma_k\Delta t})} \boldsymbol{\xi}_i^{(k)}.
\end{split}\label{eq:prop-thermo}
\end{equation}
Next, the momenta are evolved under the action of the pressure term and the inter-atomic forces
for half a time step (Eqs.~\eqref{eq:motion-force} and~\eqref{eq:motion-alpha}):
\begin{equation}
\begin{split}
\alpha \,\leftarrow\, & \alpha+ 3n \frac{\Delta t}{2} \left[V \left(P_{\rm int}-P_{\rm ext}\right)+\frac{1}{\beta}\right]  
\\
+ & \left(\frac{\Delta t}{2}\right)^2  \sum_i \frac{1}{m_i} \tilde{\mathbf{f}}_i^{(0)} \cdot \tilde{\mathbf{p}}_i^{(0)}
+ \left(\frac{\Delta t}{2}\right)^3  \sum_i \frac{1}{3 m_i} \tilde{\mathbf{f}}_i^{(0)} \cdot \tilde{\mathbf{f}}_i^{(0)}
\\
\tilde{\mathbf{p}}_i^{(k)} \,\leftarrow\, &
   \tilde{\mathbf{p}}_i^{(k)} + \tilde{\mathbf{f}}_i^{(k)} {\Delta t}/{2}.
\end{split}\label{eq:prop-pbaro}
\end{equation}
This brings us to the central term in Eq.~\eqref{eq:bigliouville}, which comprises two 
independent (and commuting) Liouvillians. One of these
(Eqs.~\eqref{eq:liouville-qscale}, \eqref{eq:liouville-pscale},
 \eqref{eq:liouville-volume} and the $k=0$ term in~\eqref{eq:liouville-nm}) 
 evolves the centroid using an algorithm identical to the stochastic classical 
 barostat of Ref.~\cite{buss+09jcp}, 
\begin{equation}
\begin{split}
\tilde{\mathbf{p}}_i^{(0)} \,\leftarrow\, &
   \tilde{\mathbf{p}}_i^{(0)} e^{-\Delta t\, \alpha/\mu} \\
\tilde{\mathbf{q}}_i^{(0)} \,\leftarrow\, &
 \tilde{\mathbf{q}}_i^{(0)} e^{\Delta t\, \alpha/\mu} +\frac{\sinh \Delta t\,\alpha/\mu}{\alpha/\mu} \frac{\tilde{\mathbf{p}}_i^{(0)}}{m_i}
\\
V \,\leftarrow\,& V e^{3\Delta t\, \alpha/\mu}, 
\end{split}
\end{equation}
while the other (comprising the terms with $k>0$ in Eq.~\eqref{eq:liouville-nm}) evolves the internal
modes using an algorithm identical to the free ring polymer 
{\em NVE} integrator in Ref.~\cite{ceri+10jcp}
\begin{equation}
\begin{pmatrix}
\tilde{\mathbf{p}}_{i}^{(k)} \cr \tilde{\mathbf{q}}_{i}^{(k)}
\end{pmatrix} 
\,\leftarrow\,
\begin{pmatrix}
\cos \omega_k\Delta t & -m_i\omega_k\sin \omega_k\Delta t \cr
[1/m_i\omega_k]\sin \omega_k\Delta t & \cos \omega_k\Delta t \cr
\end{pmatrix}
\begin{pmatrix}
\tilde{\mathbf{p}}_{i}^{(k)} \cr \tilde{\mathbf{q}}_{i}^{(k)}
\end{pmatrix}, 
\end{equation}
where $\omega_k=2\omega_n \sin k\pi/n$.
At this point, one continues with the second part of the integrator,
executing again the steps in Eqs.~\eqref{eq:prop-pbaro}, and finally applying the 
thermostats as in Eq.~\eqref{eq:prop-thermo}. Note that 
the modular nature of this integration scheme means it is very easy to implement a 
different thermostatting approach for the ions or the cell, simply by changing Eq.~\eqref{eq:prop-thermo}
to the appropriate propagator.

\section{An example application: high pressure water}

To highlight some of the more advanced features available in \ipi{}, 
we have used it to perform an illustrative \emph{ab initio} 
PIGLET \cite{ceri-mano12prl} simulation of supercritical water at 750~K and 10~GPa. 
This simulation is both technically challenging and physically interesting. 
We shall use it to reveal the importance of NQEs 
under conditions similar to those explored in the pioneering work of Ref.~\cite{schw+01prl},
in which calculations were performed at constant volume and without quantum effects
in the nuclear motion.

\subsection{Computational details}

All the present calculations used a simulation box containing 64 water molecules, initially equilibrated
at the desired state point using an empirical forcefield~\cite{habe+09jcp}. 
Starting from this configuration, we performed 25~ps of {\em NVT} dynamics with classical nuclei 
and \emph{ab initio} forces.  We employed a patched version 
of CP2K \cite{CP2K} as the electronic structure client, with computational details 
similar to those used in Ref.~\cite{schm+09jpcb}. We used the BLYP exchange-correlation 
functional \cite{beck88pra,lee+88prb} and GTH pseudopotentials \cite{goed+96prb}. 
Wave functions were expanded in the Gaussian DZVP basis set, 
while the electronic density was represented using an auxiliary plane 
wave basis, with a kinetic energy cutoff of 300~Ry.
In our constant-pressure simulations we used a TZV2P basis set, a cutoff of 800~Ry, and a plane wave
grid consistent with a cubic reference cell with a side of 10.5~\AA\ (the higher cutoff was necessary
to converge the stress tensor used to define the pressure virial).  
We included D3 empirical Van der Waals corrections~\cite{grim+10jcp,guid+09jctc,guid+10jctc},
which are essential to obtain a reasonable description of water at constant pressure. 

After the preliminary {\em NVT} equilibration, we performed a reference 50~ps {\em NPT} trajectory 
with classical nuclei, with the first 5~ps discarded for equilibration. The integration time step was 0.5~fs.
We used a stochastic velocity rescaling thermostat~\cite{buss+07jcp} for the ions, with a 
time constant of 20~fs, and an optimal-sampling GLE thermostat~\cite{ceri+10jctc} for the 
cell momentum -- for which we chose a fictitious mass consistent with $\tau_\alpha=500$~fs.  
We then performed 50~ps of simulation with quantum ions, 
modelling the quantum nature of nuclei using a 4-bead path integral
supplemented with PIGLET colored noise. The cell momentum was thermostatted using 
an optimal-sampling, classical GLE thermostat.
Input files for both \ipi{} and CP2K are provided among the examples attached to the source
code of \ipi{}.

\begin{figure}[hpbt]
\centering\includegraphics[width=\onecol]{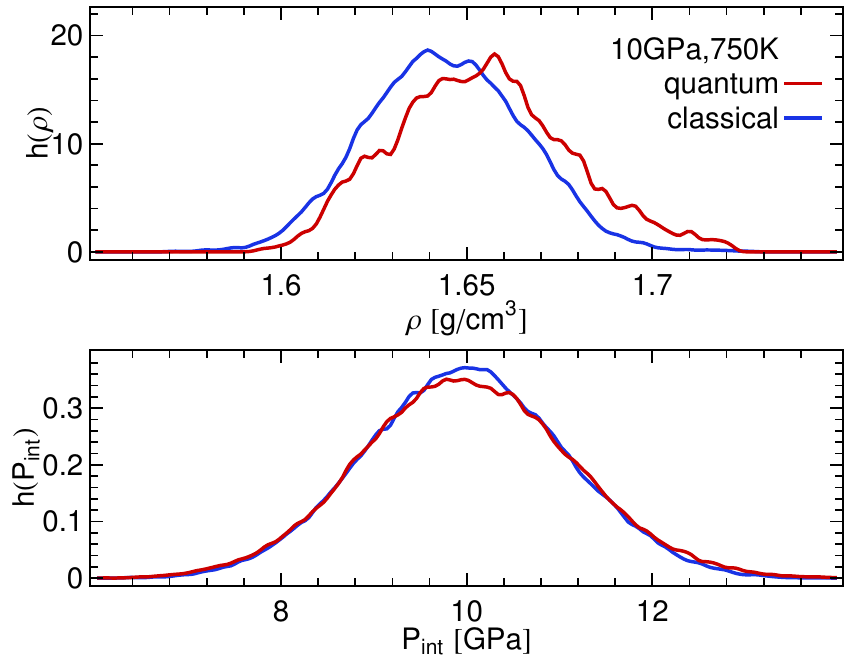}
\caption{\label{fig:histos} 
Histogram of the density (upper panel) and of the instantaneous
internal pressure estimator in Eq.~\eqref{eq:pressure} (lower panel), comparing 
{\em NPT} simulations with classical and quantum nuclei.}
\end{figure}

\subsection{Results and discussion}

Figure~\ref{fig:histos} reports the histograms of density and internal pressure obtained 
during the simulations. The average density in the classical simulation is $1.644\pm0.001$~g/cm$^3$, 
which is qualitatively consistent with the simulation of Ref.~\cite{schw+01prl}, which 
observed an average pressure of 14.5~GPa in the {\em NVT} 
ensemble at a  density of $1.72$~g/cm$^3$. 
The density appears to be slightly higher for simulations with quantum nuclei, 
$1.654\pm0.003$~g/cm$^3$, but one should note that the correlation time for $\rho$ tends to 
be underestimated in relatively short \emph{ab initio} runs, so 
the quantum effect is barely significant given that
our computed error bars are likely to be optimistic.
Note that the estimator for the internal pressure
is correctly peaked around the value fixed by the definition of the ensemble.

\begin{figure}[hpbt]
\centering\includegraphics[width=\onecol]{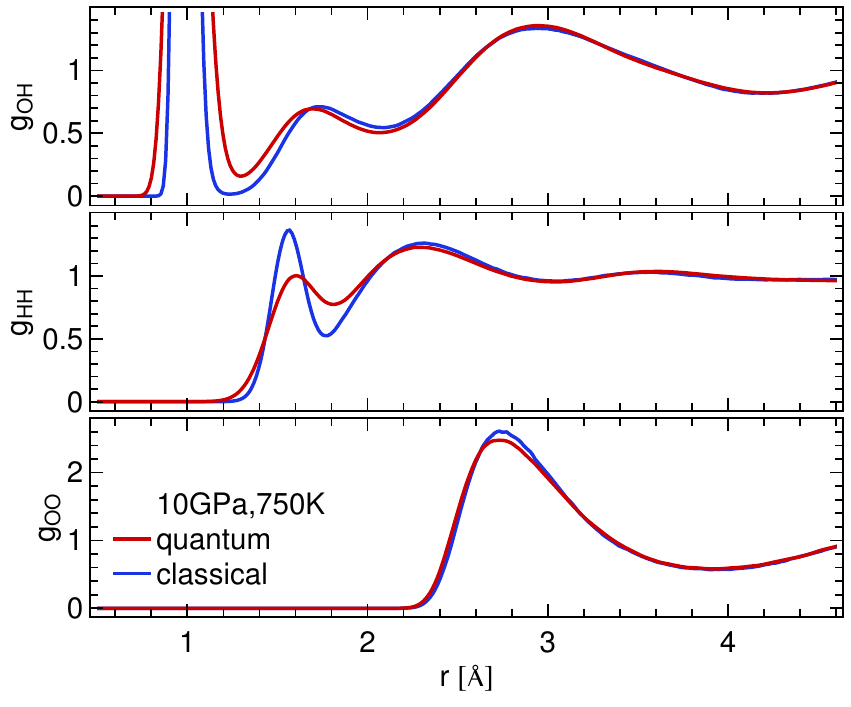}
\caption{\label{fig:gr-npt} Radial distribution functions from {\em NPT} simulations
of water at a temperature of 750~K and a pressure of 10~GPa. Simulations
with classical and quantum nuclei are compared. }
\end{figure}

The small role that NQEs play in determining the equilibrium density is not surprising: 
even at room temperature, the isotope effect on the number density of water 
is extremely small. However, one 
should not think that at these high temperatures the quantum nature of light nuclei can be 
neglected. It is clear from the radial distribution functions reported in Figure~\ref{fig:gr-npt}
that, while nuclear quantum effects do not change the long-range structure, they do have
a very sizeable impact on the short-range part of $g(r)$. There is very little effect
on the oxygen-oxygen $g_\text{OO}(r)$ (the difference is barely significant considering the statistical 
error bars), but  $g_\text{HH}(r)$ and $g_\text{OH}(r)$ are noticeably over-structured
for $r<2\text{ \AA}$ in the classical simulation. Notice
in particular how the quantum $g_\text{OH}(r)$ departs significantly from zero
between the intra-molecular and inter-molecular regions. This is a sign that the combined 
effect of pressure and NQEs leads to significant delocalization of protons along the hydrogen 
bonds.

\begin{figure}[hpbt]
\centering\includegraphics[width=\onecol]{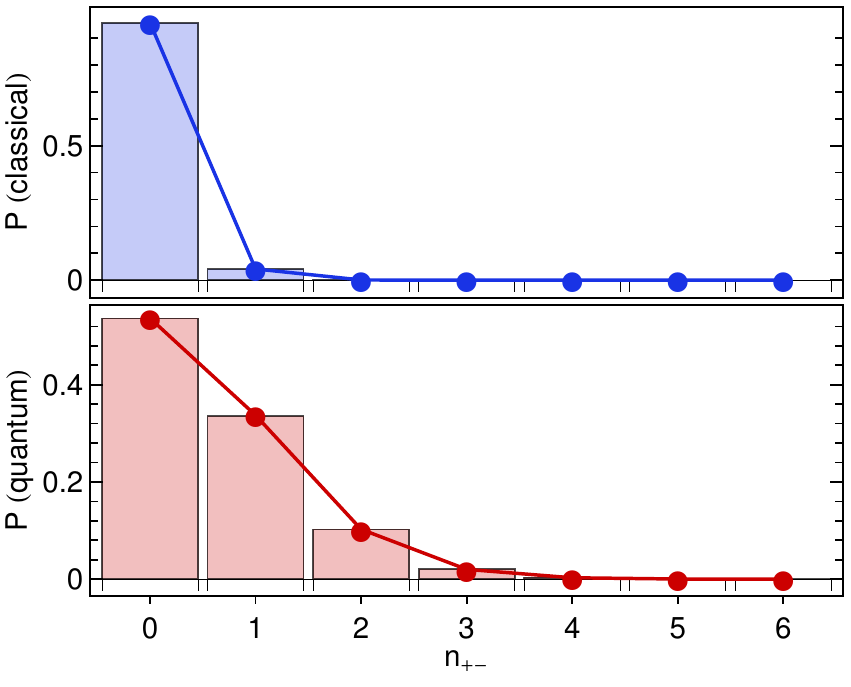}
\caption{\label{fig:barchart} These bar charts depict the probability of finding 
$n_{+-}$ charged pairs in a snapshot of our simulation box. 
The upper panel is for a classical simulation, and the lower panel
for a quantum simulation. The dots joined with lines represent the values for a 
binomial distribution fitted to the computed probabilities -- 
each snapshot contains 64 water molecules, so can accommodate
a maximum of 32 ion pairs. The best-fit probability of a pair being ionized is 
found to be $p_{+-}=0.019$ for the quantum simulation and 
$p_{+-}=0.0014$ for the classical simulation.}
\end{figure}

To investigate this delocalisation further, we have adopted the simple protocol
used in Ref.~\cite{schw+01prl} to identify charged species. 
Each proton in the simulation is assigned to the closest oxygen atom.
One can then distinguish between neutral, positively and negatively charged
species based on the number of protons assigned to each oxygen.
We will refer to ``$+$'' species as those oxygen atoms to which three protons 
have been assigned, and ``$-$'' species as those with just one. 
Given the artificial nature of this procedure, we do not intend to
imply that these species correspond to water, hydronium and hydroxide. However, 
they do provide a simple
way to characterise how much each frame departs from the conventional picture of a molecular
fluid composed of neutral molecules. In all of our simulations, we only detected a single instance
of a ``doubly ionised'' oxygen, so in the discussion that follows we shall assume that the
charged O atoms always form in pairs.

Figure~\ref{fig:barchart} shows very clearly just how important NQEs 
are in determining the fluctuations of  protons along hydrogen bonds. 
Most frames in the classical simulation only contain neutral
species -- the concentration of $+$ or $-$ is $\approx 0.07$\%, even smaller than that
observed in Ref.~\cite{schw+01prl}, where the density was higher. 
In the quantum simulation, the concentration
is much larger, about $0.97$\%, and there is a fairly large probability of having more than one pair
of ions in any given frame. Interestingly, in both cases the probability of finding $n_{+-}$ ionised
pairs follows closely a binomial distribution, which is a sign that the ionisation events are weakly
correlated with each other. 

\begin{figure}[hpbt]
\centering\includegraphics[width=\onecol]{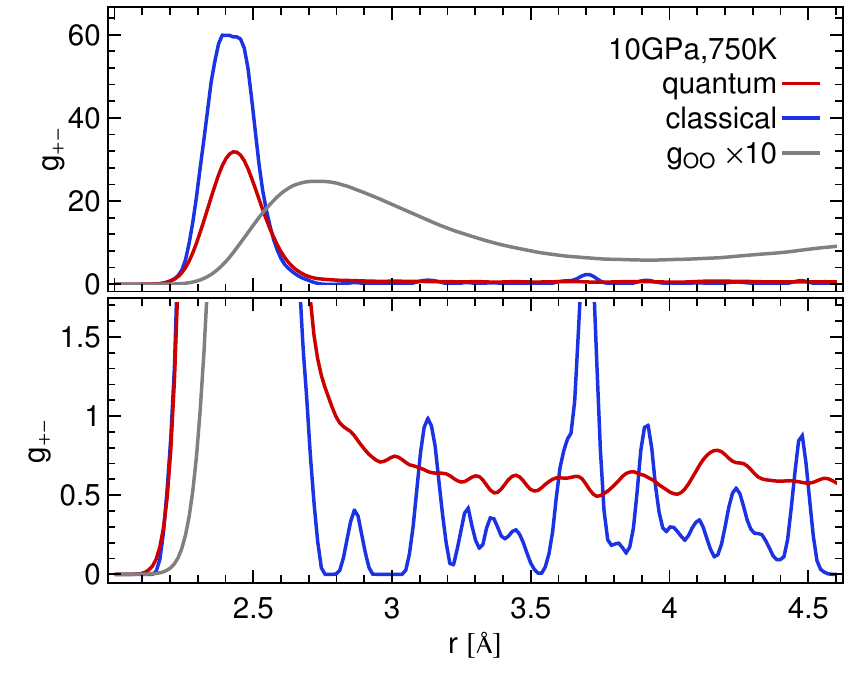}
\caption{\label{fig:gr-ions}  Radial distribution functions for positively and negatively 
charged defects in {\em NPT} simulations of water at 10~GPa and 750~K. 
Results from quantum and classical simulations 
are compared, and the lower panel shows the long-range part of $g_{+-}(r)$ on an enlarged scale.
 Given the low concentration of defects, the radial distribution functions are very noisy, and
the normalisation is problematic, so the scale of the ordinate should be considered as arbitrary.}
\end{figure}
 
It is interesting to investigate the spatial distribution of these ion pairs: in Figure~\ref{fig:gr-ions}
we show the radial distribution functions $g_{+-}(r)$.
The distributions are very noisy, particularly in the classical case, where we have
only a handful of snapshots containing more than a single ion pair. 
In the quantum simulation,  we could collect better statistics, 
because the concentration of ion pairs is larger. 
In this case, is clear that the radial distribution function is almost flat for $r>3$~\AA{} 
(see lower panel of Figure~\ref{fig:gr-ions}), 
which is consistent with the binomial distribution of $n_{+-}$ in Figure~\ref{fig:barchart}.   
Furthermore, the upper panel of Figure~\ref{fig:gr-ions}  clearly shows that,
 in the vast majority of cases, what is detected as a pair of ions
is simply an excursion of a proton along a compressed hydrogen bond. This is consistent with 
the observation of short-lived charged species in classical simulations of water at considerably
higher pressure~\cite{gonc+05prl}. 

As was shown in a recent study of  
water under ambient conditions~\cite{pnasinpress}, nuclear quantum 
effects  dramatically enhance the delocalisation of the proton along the 
hydrogen bond. This suggests that perturbations 
that modulate the average O--O distance, such as pressure, 
might trigger auto-ionisation more easily than in the classical case. 
In fact, not all of the ion pairs in the present simulations are close 
together and short-lived: if we define an isolated ion
as a positive species that has no negative counterpart within $3$~\AA{} of it, we find a concentration 
of $6\times 10^{-3}$\%~of these isolated ions in our classical simulations, and of $0.24$\%~in 
our quantum simulations.
This is not far from the value of 0.5~\%~measured in shock-compressed
water at 13~GPa~\cite{hama74book,mitc-nell82jcp}.
However, our definition of an isolated ion is somewhat arbitrary, 
and the concentrations we quote will be strongly system-size dependent. 
Nevertheless, the ratio between these concentrations and the overall fraction of ionised species
indicates that quantum effects have an even more dramatic impact on 
genuine auto-ionisation events than they do on local hydrogen bond  fluctuations. 
This is perhaps not too surprising, given the significant isotope effect on the 
pH of water under ambient conditions. 

\begin{figure}[hpbt]
\centering\includegraphics[width=\onecol]{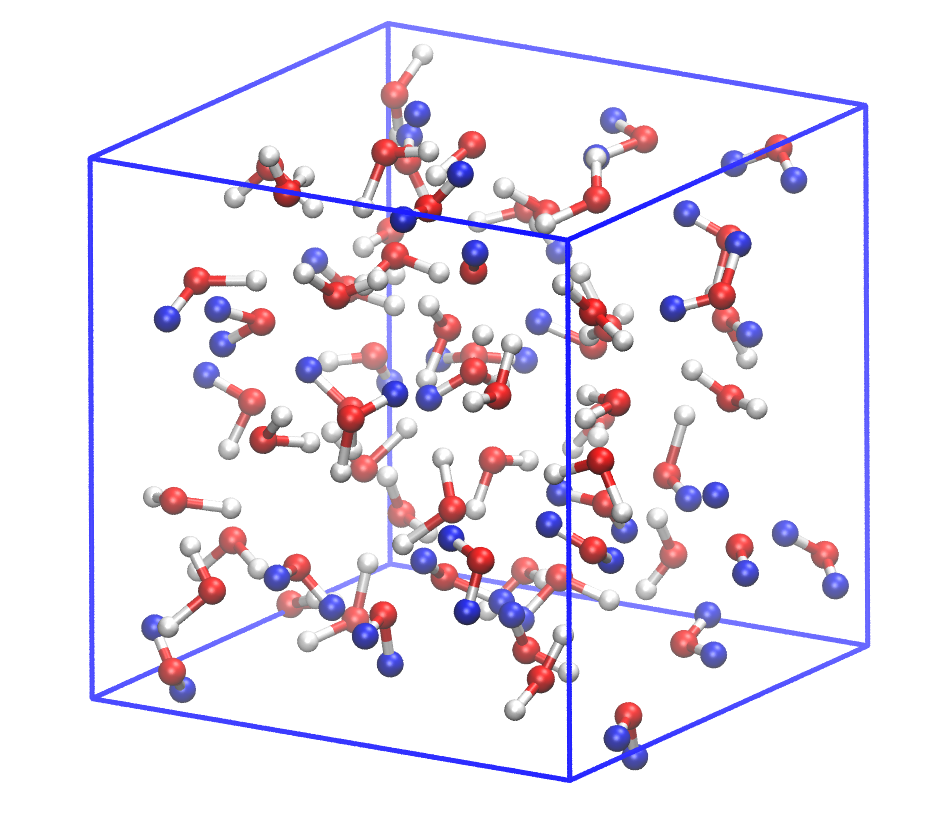}
\caption{\label{fig:colors}  A snapshot from the final stage of our 40~ps 
 quantum simulation of water at 750~K and 10~GPa. 
 Protons that are bound to a different oxygen than the one they
were bound to in the initial configuration of the trajectory are highlighted in blue.
Note that many exchanges have occurred. In our classical simulation, no exchanges 
were observed. }
\end{figure}

The presence of a significant fraction of ionised species increases the mobility of protons, which
can hop from molecule to molecule in a Grotthuss-like fashion. Even though PIMD (and particularly
the heavily-thermostatted PIGLET method) does not allow us to make quantitative statements 
about how quantum effects enhance proton mobility, one can clearly see that 
by the end of our quantum simulation a significant fraction of the protons
has been exchanged between water molecules (see Figure~\ref{fig:colors}).
On the contrary, in the classical case all the water molecules
maintain their chemical identity thoughout our simulation.
Using an approximate quantum dynamics technique
such as RPMD~\cite{habe+13arpc} or CMD~\cite{cao-voth93jcp,cao-voth94jcp} 
to study the dynamics in high pressure 
water would be an interesting future application of~\ipi{}.

\section{Conclusions}

In this paper we have introduced \ipi{}, a Python interface designed to facilitate including
nuclear quantum effects in \emph{ab initio} path integral molecular dynamics simulations.
Our program delegates the calculation of the potential, forces and virial
tensor to an external code, keeping the electronic structure calculation
and the propagation of the nuclear dynamics separate.
Communication between the codes is achieved using internet sockets, which exchange
just the essential  information, thereby
minimising the number of modifications that have to be made to the \emph{ab initio} client.
\ipi{} implements the most recent developments in path integral simulation
technology, including correlated-noise methods that accelerate the  convergence
with respect to the number of beads, and an implementation of {\em NPT} PIMD
based on stochastic thermostatting.

We have demonstrated the potential of \ipi{} with an application to the
\emph{ab initio} simulation of high-pressure water. Our simulations with classical 
nuclei are consistent with previous simulations~\cite{schw+01prl,gonc+05prl}, 
performed at a similar thermodynamic state point
and using similar computational details for the electronic structure. However,
our results show that even at a temperature as high as 750~K nuclear quantum effects 
play an important role in determining the behaviour of water. The concentration
of ionised species -- as defined by a simple geometric criterion -- is increased by more than
an order of magnitude when one treats the nuclei as quantum particles. 
We have characterised the nature of these charged species, observing that in the vast majority
of cases they correspond to fluctuations of a proton along a compressed hydrogen bond,
giving rise to a transient contact pair rather than to well-separated, solvated charges. 
Nevertheless, at the high ionic concentrations (more than 1~\%) observed in quantum simulations,
a smaller fraction of these fluctuations leads to long-range separation of ionised 
species, and to effective transport and exchange of protons across the hydrogen-bond 
network, which is observed in classical simulations only at a 
much higher pressure~\cite{schw+01prl,gonc+05prl}. 

These results and those of many other recent studies are leading to an increasing body of 
evidence that it is desirable (and in some cases even essential) to include nuclear quantum 
effects in molecular dynamics simulations if one wants to obtain a realistic description of systems 
containing hydrogen atoms. The inclusion of quantum effects certainly entails a computational 
overhead, but recent methodological developments that combine path integrals with correlated-noise, 
generalized Langevin dynamics have reduced this overhead considerably.
We envisage that the introduction of \ipi{} will make these new techniques
more readily accessible, and thereby encourage the more routine inclusion of NQEs
in \emph{ab initio} molecular dynamics simulations. The modular nature of \ipi{} should also ensure that
further methodological developments will be easy to incorporate,
so that communities that use a diverse variety of electronic structure 
programs can be kept up-to-date with the field.

\section*{Acknowledgements}

We would like to thank the early adopters of \ipi{}, 
including  J. Cuny, E. Davidson, R. DiStasio, D. Donadio, F. Giberti, A. Hassanali, 
T. Markland, M. Rossi, B. Santra, D. Selassie, T. Spura, L. Wang, and D. Wilkins, 
who have helped immensely debugging the code and testing it on a variety of problems.
A special thanks goes to A. Michaelides, who helped us to decide an acronym for the interface. 
We also acknowledge generous allocations of computer time from CSCS (project ID s338) 
and the Oxford Supercomputing Centre, and funding from the EU Marie Curie 
IEF No. PIEF- GA-2010-272402, the Wolfson Foundation and the Royal Society.





\bibliographystyle{elsarticle-num}

\begin{thebibliography}{10}
\expandafter\ifx\csname url\endcsname\relax
  \def\url#1{\texttt{#1}}\fi
\expandafter\ifx\csname urlprefix\endcsname\relax\def\urlprefix{URL }\fi
\expandafter\ifx\csname href\endcsname\relax
  \def\href#1#2{#2} \def\path#1{#1}\fi

\bibitem{fisc+06natm}
C.~C. Fischer, K.~J. Tibbetts, D.~Morgan, G.~Ceder, {Predicting crystal
  structure by merging data mining with quantum mechanics.}, Nature materials
  5~(8) (2006) 641--6.

\bibitem{car-parr85prl}
R.~Car, M.~Parrinello, {Unified Approach for Molecular Dynamics and
  Density-Functional Theory}, Phys. Rev. Lett. 55~(22) (1985) 2471--2474.

\bibitem{burk12jcp}
K.~Burke, {Perspective on density functional theory.}, J. Chem. Phys. 136~(15)
  (2012) 150901.

\bibitem{chan-woly81jcp}
D.~Chandler, P.~G. Wolynes, {Exploiting the isomorphism between quantum theory
  and classical statistical mechanics of polyatomic fluids}, J. Chem. Phys.
  74~(7) (1981) 4078--4095.

\bibitem{parr-rahm84jcp}
M.~Parrinello, A.~Rahman, {Study of an F center in molten KCl}, J. Chem. Phys.
  80 (1984) 860.

\bibitem{cepe95rmp}
D.~M. Ceperley, {Path integrals in the theory of condensed helium}, Rev. Mod.
  Phys. 67~(2) (1995) 279--355.

\bibitem{feyn-hibb65book}
R.~P. Feynman, A.~R. Hibbs, {Quantum Mechanics and Path Integrals},
  McGraw-Hill, New York, 1964.

\bibitem{tuck+96jcp}
M.~E. Tuckerman, D.~Marx, M.~L. Klein, M.~Parrinello, {Efficient and general
  algorithms for path integral Car\{--\}Parrinello molecular dynamics}, J.
  Chem. Phys. 104~(14) (1996) 5579--5588.

\bibitem{marx-parr96jcp}
D.~Marx, M.~Parrinello, {Ab initio path integral molecular dynamics: Basic
  ideas}, J. Chem. Phys. 104~(11) (1996) 4077--4082.

\bibitem{marx+99nat}
D.~Marx, M.~E. Tuckerman, J.~Hutter, M.~Parrinello, {The nature of the hydrated
  excess proton in water}, Nature 397~(6720) (1999) 601--604.

\bibitem{haye+09jpcb}
R.~L. Hayes, S.~J. Paddison, M.~E. Tuckerman, {Proton transport in triflic acid
  hydrates studied via path integral car-parrinello molecular dynamics.}, J.
  Phys. Chem.. B 113~(52) (2009) 16574--89.

\bibitem{chen+13pccp}
J.~Chen, X.-Z. Li, Q.~Zhang, A.~Michaelides, E.~Wang, {Nature of proton
  transport in a water-filled carbon nanotube and in liquid water.}, PCCP
  (2013) 6344--6349.

\bibitem{buyu+08pre}
S.~Buyukdagli, A.~V. Savin, B.~Hu, {Computation of the temperature dependence
  of the heat capacity of complex molecular systems using random color noise},
  Phys. Rev. E 78~(6) (2008) 66702.

\bibitem{ceri+09prl2}
M.~Ceriotti, G.~Bussi, M.~Parrinello, {Nuclear quantum effects in solids using
  a colored-noise thermostat}, Phys. Rev. Lett. 103~(3) (2009) 30603.

\bibitem{damm+09prl}
H.~Dammak, Y.~Chalopin, M.~Laroche, M.~Hayoun, J.-J. Greffet, {Quantum Thermal
  Bath for Molecular Dynamics Simulation}, Phys. Rev. Lett. 103~(19) (2009)
  190601.

\bibitem{ceri+11jcp}
M.~Ceriotti, D.~E. Manolopoulos, M.~Parrinello, {Accelerating the convergence
  of path integral dynamics with a generalized Langevin equation.}, J. Chem.
  Phys. 134~(8) (2011) 84104.

\bibitem{ceri-mano12prl}
M.~Ceriotti, D.~E. Manolopoulos, {Efficient First-Principles Calculation of the
  Quantum Kinetic Energy and Momentum Distribution of Nuclei}, Phys. Rev. Lett.
  109~(10) (2012) 100604.

\bibitem{ceri+10jcp}
M.~Ceriotti, M.~Parrinello, T.~E. Markland, D.~E. Manolopoulos, {Efficient
  stochastic thermostatting of path integral molecular dynamics.}, J. Chem.
  Phys. 133~(12) (2010) 124104.

\bibitem{tuck+92jcp}
M.~Tuckerman, B.~J. Berne, G.~J. Martyna, {Reversible multiple time scale
  molecular dynamics}, J. Chem. Phys. 97~(3) (1992) 1990.

\bibitem{mark-mano08jcp}
T.~E. Markland, D.~E. Manolopoulos, {An efficient ring polymer contraction
  scheme for imaginary time path integral simulations.}, J. Chem. Phys. 129~(2)
  (2008) 024105.

\bibitem{mark-mano08cpl}
T.~E. Markland, D.~E. Manolopoulos, {A refined ring polymer contraction scheme
  for systems with electrostatic interactions}, Chem. Phys. Lett. 464~(4-6)
  (2008) 256.

\bibitem{buss+07jcp}
G.~Bussi, D.~Donadio, M.~Parrinello, {Canonical sampling through velocity
  rescaling}, J. Chem. Phys. 126~(1) (2007) 14101.

\bibitem{ceri+09prl}
M.~Ceriotti, G.~Bussi, M.~Parrinello, {Langevin Equation with Colored Noise for
  Constant-Temperature Molecular Dynamics Simulations}, Phys. Rev. Lett.
  102~(2) (2009) 020601.

\bibitem{ceri+10jctc}
M.~Ceriotti, G.~Bussi, M.~Parrinello, {Colored-Noise Thermostats \`{a} la
  Carte}, J. Chem. Theory Comput. 6~(4) (2010) 1170--1180.

\bibitem{ceri-parr10pcs}
M.~Ceriotti, M.~Parrinello, {The -thermostat: selective normal-modes excitation
  by colored-noise Langevin dynamics}, Procedia Computer Science 1~(1) (2010)
  1607--1614.

\bibitem{gle4md}
{GLE4MD}, http://gle4md.berlios.de.

\bibitem{yama05jcp}
T.~M. Yamamoto, {Path-integral virial estimator based on the scaling of
  fluctuation coordinates: Application to quantum clusters with fourth-order
  propagators}, J. Chem. Phys. 123~(10) (2005) 104101.

\bibitem{ceri-mark13jcp}
M.~Ceriotti, T.~E. Markland, {Efficient methods and practical guidelines for
  simulating isotope effects.}, J. Chem. Phys. 138~(1) (2013) 014112.

\bibitem{lin+10prl}
L.~Lin, J.~A. Morrone, R.~Car, M.~Parrinello, {Displaced Path Integral
  Formulation for the Momentum Distribution of Quantum Particles}, Phys. Rev.
  Lett. 105~(11) (2010) 110602.

\bibitem{crai-mano04jcp}
I.~R. Craig, D.~E. Manolopoulos, {Quantum statistics and classical mechanics:
  Real time correlation functions from ring polymer molecular dynamics}, J.
  Chem. Phys. 121 (2004) 3368.

\bibitem{habe+13arpc}
S.~Habershon, D.~E. Manolopoulos, T.~E. Markland, T.~F. Miller, {Ring-polymer
  molecular dynamics: quantum effects in chemical dynamics from classical
  trajectories in an extended phase space.}, Annual review of physical
  chemistry 64 (2013) 387--413.

\bibitem{cao-voth93jcp}
J.~Cao, G.~A. Voth, {A new perspective on quantum time correlation functions},
  J. Chem. Phys. 99~(12) (1993) 10070--10073.

\bibitem{cao-voth94jcp}
J.~Cao, G.~A. Voth, {The formulation of quantum statistical mechanics based on
  the Feynman path centroid density. IV. Algorithms for centroid molecular
  dynamics}, J. Chem. Phys. 101~(7) (1994) 6168--6183.

\bibitem{buss+09jcp}
G.~Bussi, T.~Zykova-Timan, M.~Parrinello, {Isothermal-isobaric molecular
  dynamics using stochastic velocity rescaling}, J. Chem. Phys. 130~(7) (2009)
  74101.

\bibitem{mart+99jcp}
G.~J. Martyna, A.~Hughes, M.~E. Tuckerman, {Molecular dynamics algorithms for
  path integrals at constant pressure}, J. Chem. Phys. 110~(7) (1999) 3275.

\bibitem{ande80jcp}
H.~C. Andersen, {Molecular dynamics simulations at constant pressure and/or
  temperature}, J. Chem. Phys. 72~(4) (1980) 2384--2393.

\bibitem{parr-rahm81jap}
M.~Parrinello, {Polymorphic transitions in single crystals: A new molecular
  dynamics method}, J. Appl. Phys. 52~(12) (1981) 7182.

\bibitem{buss-parr07pre}
G.~Bussi, M.~Parrinello, {Accurate sampling using Langevin dynamics}, Phys.
  Rev. E 75~(5) (2007) 56707.

\bibitem{schw+01prl}
E.~Schwegler, G.~Galli, F.~Gygi, R.~Hood, {Dissociation of Water under
  Pressure}, Phys. Rev. Lett. 87~(26) (2001) 265501.

\bibitem{habe+09jcp}
S.~Habershon, T.~E. Markland, D.~E. Manolopoulos, {Competing quantum effects in
  the dynamics of a flexible water model.}, J. Chem. Phys. 131~(2) (2009)
  24501.

\bibitem{CP2K}
{\{CP2K\}}, http://cp2k.berlios.de.

\bibitem{schm+09jpcb}
J.~Schmidt, J.~VandeVondele, I.-F.~W. Kuo, D.~Sebastiani, J.~I. Siepmann,
  J.~Hutter, C.~J. Mundy, {Isobaric-isothermal molecular dynamics simulations
  utilizing density functional theory: an assessment of the structure and
  density of water at near-ambient conditions.}, J. Phys. Chem.. B 113~(35)
  (2009) 11959--64.

\bibitem{beck88pra}
A.~D. Becke, {Density-functional exchange-energy approximation with correct
  asymptotic behavior}, Phys. Rev. A 38~(6) (1988) 3098.

\bibitem{lee+88prb}
C.~Lee, W.~Yang, R.~G. Parr, {Development of the Colle-Salvetti
  correlation-energy formula into a functional of the electron density}, Phys.
  Rev. B 37~(2) (1988) 785.

\bibitem{goed+96prb}
S.~Goedecker, M.~Teter, J.~Hutter, {Separable dual-space Gaussian
  pseudopotentials}, Phys. Rev. B 54~(3) (1996) 1703--1710.

\bibitem{grim+10jcp}
S.~Grimme, J.~Antony, S.~Ehrlich, H.~Krieg, {A consistent and accurate ab
  initio parametrization of density functional dispersion correction (DFT-D)
  for the 94 elements H-Pu.}, J. Chem. Phys. 132~(15) (2010) 154104.

\bibitem{guid+09jctc}
M.~Guidon, J.~Hutter, J.~VandeVondele, {Robust Periodic Hartree−Fock Exchange
  for Large-Scale Simulations Using Gaussian Basis Sets}, J. Chem. Theory
  Comput. 5~(11) (2009) 3010--3021.

\bibitem{guid+10jctc}
M.~Guidon, J.~Hutter, J.~VandeVondele, {Auxiliary Density Matrix Methods for
  Hartree−Fock Exchange Calculations}, J. Chem. Theory Comput. 6~(8) (2010)
  2348--2364.

\bibitem{gonc+05prl}
A.~Goncharov, N.~Goldman, L.~Fried, J.~Crowhurst, I.-F. Kuo, C.~Mundy, J.~Zaug,
  {Dynamic Ionization of Water under Extreme Conditions}, Phys. Rev. Lett.
  94~(12) (2005) 125508.

\bibitem{pnasinpress}
M.~Ceriotti, J.~Cuny, M.~Parrinello, D.~E. Manolopoulos, Nuclear quantum
  effects and hydrogen bond fluctuations in water, Proc. Nat. Acad. Sci.
  110~(39) (2013) 15591--15596.

\bibitem{hama74book}
S.~D. Hamann, {Electrolyte solutions at high pressure}, in: Modern aspects of
  electrochemistry, Springer, 1974, pp. 47--158.

\bibitem{mitc-nell82jcp}
A.~C. Mitchell, W.~J. Nellis, {Equation of state and electrical conductivity of
  water and ammonia shocked to the 100 GPa (1 Mbar) pressure range}, J. Chem.
  Phys. 76~(12) (1982) 6273.

\end{thebibliography}

\end{document}